\documentclass[a4paper, amsfonts, amssymb, amsmath, preprint, showkeys, nofootinbib, twoside]{revtex4-1}
\usepackage[english]{babel}
\usepackage[utf8]{inputenc}
\usepackage[colorinlistoftodos, color=green!40, prependcaption]{todonotes}
\usepackage{amsthm}
\usepackage{mathtools}
\usepackage{physics}
\usepackage{xcolor}
\usepackage{graphicx}
\usepackage[left=23mm,right=13mm,top=35mm,columnsep=15pt]{geometry} 
\usepackage{adjustbox}
\usepackage{placeins}
\usepackage[T1]{fontenc}
\usepackage{lipsum}
\usepackage{csquotes}
\usepackage[pdftex, pdftitle={Article}, pdfauthor={Author}]{hyperref} 
\begin{document}

\title{
Win Probabilities, Hand Sizes, and Game Duration Analysis in
the \textit{Bhikar-Sawkar} Card Game}

\author{Mihir Durve}
    \email[Correspondence email address: ]{mihir.durve@iit.it}
    \affiliation{Center for Life Nano- \& Neuro-Science, Italian Institute of Technology (IIT), viale Regina Elena 295, Rome, 00161, Italy}

\date{\today} 

\begin{abstract}
We present a Monte Carlo simulation study of the \textit{Bhikar-Sawkar} card game, a non-deterministic game structurally similar to the classic \textit{Beggar-My-Neighbour}, which is fully deterministic. Although both games share a common setup, key differences in their rules—particularly the reshuffling of cards after each won hand in \textit{Bhikar-Sawkar}—introduce stochasticity and significantly increase the space of possible game evolutions. This inherent randomness raises a range of interesting statistical questions regarding the duration of the game, the hand-winner distributions, and the probability of winning the game for a given player. These questions are systematically investigated through large-scale simulations across multiple game configurations.
  
\end{abstract}

\keywords{Monte Carlo method, Probabilities in card games, Bhikar-Sawkar, Beggar-My-Neighbour}

\maketitle

\section{Introduction} \label{sec:outline}

Card games provide good examples for studying probability theory and combinatorics\cite{Sharna_Sharma_Doyle_Marcelo_Kumar_2021, asee_peer_27135, madsen}. While many games involve a mixture of strategic decision-making and chance, others are governed entirely by luck, offering no opportunity for player strategy. Despite their simplicity, such purely chance-based games present intriguing challenges for analyzing outcome probabilities and expected game durations. One notable example is the classic Beggar-My-Neighbour \cite{Paulhus01021999,casella2024}, a deterministic game that has drawn scholarly attention for its unpredictable behaviour and complex dynamics arising from deceptively simple rules. The classic Beggar-My-Neighbour has some non-terminating sequences, but recently, some versions of this game have been shown to always end in finite steps \cite{Lakshtanov2013}.  In this study, we examine the Bhikar-Sawkar game, a lesser-known but structurally analogous nondeterministic game, through extensive Monte Carlo simulations to study its probabilities.

Bhikar-Sawkar is a traditional card game played in Maharashtra, India, with its name originating from the Marathi language. Translated roughly to Beggar-Usurer in English, the game has been passed down through generations, often with minor variations in its rules. It accommodates a flexible number of participants, typically ranging from 2 to N players, and can be played using one or more decks of standard playing cards. Although a detailed description of the game rules is provided in Section \ref{sec:game_rules}, the fundamental objective remains consistent: the game continues until only one player remains.

Despite being governed by entirely deterministic rules, the game's complexity arises from its evolving state space through card shuffling, making analytical solutions to questions such as minimum and maximum game durations elusive. In this study, we employ Monte Carlo simulation as a computational approach to investigate these properties. By simulating millions of games across various configurations of players and card decks, we examine key statistical quantities, including the distribution of game durations, hand-winning sizes, and player win probabilities. This work contributes toward a deeper probabilistic understanding of a statistically rich, nondeterministic game.

This paper is organised as follows. We formalise the game setup and rules in the next section, Sec. \ref{sec:game_rules}. Later, we define key questions in Sec. \ref{kq} that we address in this study. The results will be discussed in the following section, Sec. \ref{sec_r_d}. At last, we note our conclusions in Sec. \ref{sec:conclusions}.

\section{Game setup} \label{sec:game_rules}

Below, we describe the game setup and flow of the \textit{Bhikar-Sawkar} game considered in this study in detail. 

\noindent 1. Setup: \\
The \textit{Bhikar-Sawkar} game is played by N players seated in a circle, facing each other. The game uses K standard decks of 52 playing cards each, which are thoroughly shuffled together to form a single combined deck. A player is randomly selected to begin the distribution (designated as Player 1), and the cards are dealt one by one in a clockwise direction, starting from this player, until the entire deck is exhausted. Cards are dealt face-down, and players are not allowed to view their own cards or those of others. Each player forms a personal, private deck from the cards they receive.

\noindent 2. Game Flow: \\
The game begins with Player 1 placing the top card of their private deck face-up in the center of the table. The act of playing a card counts as a turn. The next player in the clockwise direction (Player 2) then plays the top card of their own deck on top of the central pile. At each turn, the current player’s card is compared to the card immediately preceding it in the central pile. Two outcomes are possible:

\noindent a) Match: If the two cards share the same rank (e.g., both are 3s or both are Kings), regardless of suit, the current player is said to have won the hand. They collect all the cards in the central pile, shuffle them randomly, and append them to the bottom of their private deck. The same player then immediately plays the top card from their updated deck to continue the game.

\noindent b) No Match: If the card does not match the rank of the previous card, the turn passes to the next player clockwise, who plays the top card from their deck. This process repeats.

A player is eliminated from the game when they exhaust all the cards in their private deck. Eliminated players no longer participate in gameplay. The game proceeds in this manner until only one player remains with cards. This player is declared the winner.

\section{Key questions with \textit{Bhikar-Sawkar}}
\label{kq}
The setup of the \textit{Bhikar-Sawkar} game bears a strong resemblance to the well-studied \textit{Beggar-My-Neighbour} game. However, there are key differences in game mechanics that have significant implications for the game's dynamics. In \textit{Beggar-My-Neighbour}, when a court card (e.g., Jack, Queen, King, Ace) is played, the opposing player is required to play a specific number of response cards (ranging from one to four, depending on the court card). The player who ultimately wins the hand collects the entire central pile and appends it to the bottom of their private deck \emph{without} shuffling. This mechanism renders the game entirely deterministic and has been shown to potentially lead to non-terminating games under certain conditions.

In contrast, \textit{Bhikar-Sawkar} introduces an essential stochastic element: the player who wins a hand shuffles the collected central pile before appending it to the bottom of their deck. This reshuffling step introduces randomness into the game, making it fundamentally non-deterministic. As a consequence, while \textit{Beggar-My-Neighbour} can result in cyclic, non-terminating sequences, the reshuffling mechanism in \textit{Bhikar-Sawkar} prevents such cycles, thereby ensuring that every game eventually terminates with a single winner. This key distinction also makes the theoretical analysis of maximum game duration more elusive and reinforces the utility of Monte Carlo simulation in understanding the statistical properties of the game.

In this study, we investigate three key questions regarding the statistics of the \textit{Bhikar-Sawkar} game. First, for a given number of players and a given number of card decks, what are the expected maximum and minimum game durations in terms of number of turns before the game ends? Second, what is the empirical distribution of hand-winning sizes observed during gameplay? Third, how are the win probabilities distributed among the players?

These questions are addressed through extensive Monte Carlo simulations conducted across a wide range of game configurations. The results are discussed in the next section.

\section{Results and discussions}
\label{sec_r_d}
The following statistics are calculated over $10^6$ games for each combination of the number of players $N$ and the number of decks  $K$.  Specifically, we simulate all combinations of $N \in \{2, 3, 4, 5\}$ and $K \in \{1, 2, 3, 4, 5\}$. The results presented below summarise the observed game durations, hand-winning distributions, and win probabilities for each configuration. To support reproducibility and further exploration, the simulation code has been made publicly available at \cite{code_link}.

\subsection{Maximum and minimum game length}

\begin{figure*}
\centering
\includegraphics[width=\textwidth]{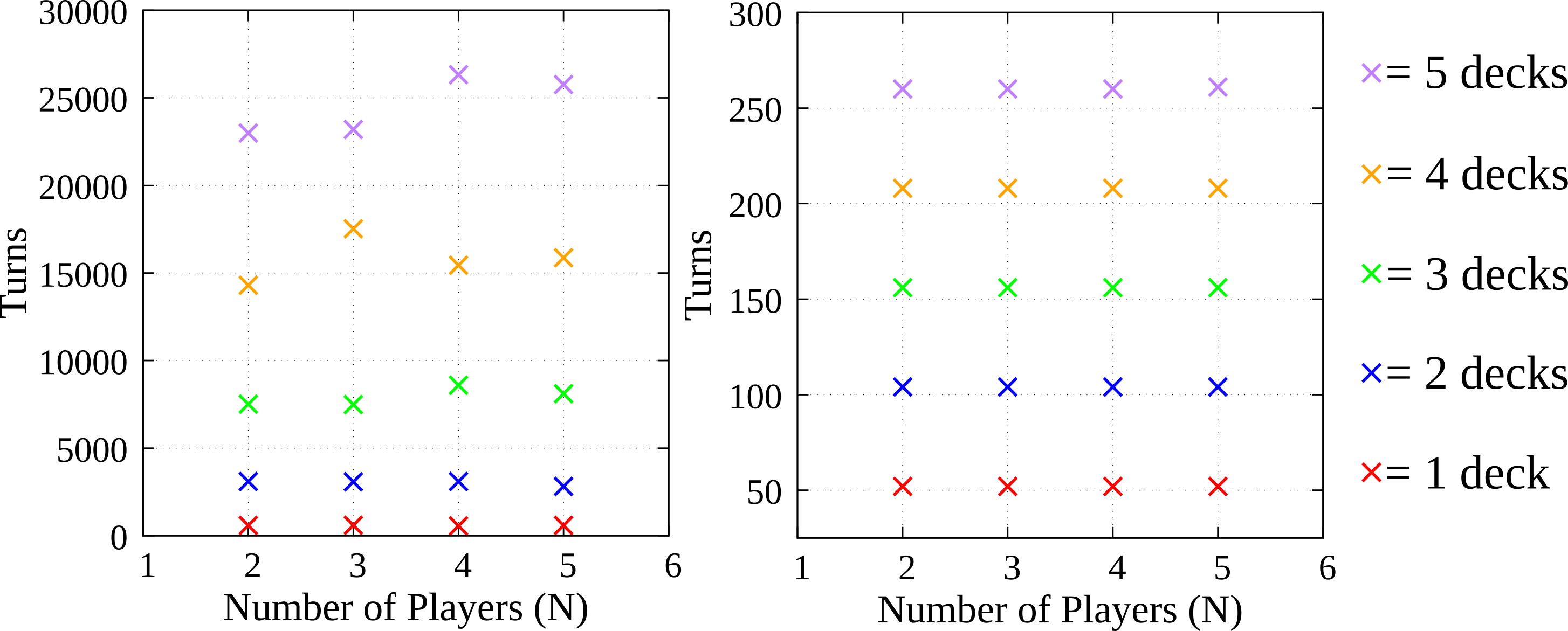}
\caption{Left panel (a) shows maximum turns and right panel (b) shows minimum number of turns in a single game. The figure is to be read with the colour code for a number of decks $K$ shown at the rightmost part of the figure. The maximum and minimum number of games are found in a simulation of $10^6$ games for each combination of number of players and number of decks. \label{fig_min_max}}
\end{figure*}

A central question in the analysis of the \textit{Bhikar-Sawkar} game concerns the possible range of game durations for a given number of players and card decks. Game duration is quantified by the total number of turns, where a turn is defined as the act of a player playing a single card. Notably, when a player wins a hand, collects the central pile, and immediately plays the next card, two turns are counted—one for the winning move and another for the subsequent play. Figure~\ref{fig_min_max} presents the observed maximum and minimum number of turns across $10^6$ simulated games for each combination of players ($N$) and decks ($K$).

As shown in Figure~\ref{fig_min_max}(b), the minimum number of turns closely corresponds to the total number of cards in games with a small number of decks. This observation is reasonable, as a smaller card pool reduces the likelihood of players winning hands, resulting in game durations that approach the total number of cards dealt. For example, in games with a single deck (52 cards), the minimum number of turns observed is 52, indicating cases where one or no players win any hands. As the number of decks increases, the probability that only a single player wins hands consistently diminishes, leading to longer minimum durations. Nevertheless, in games with fewer players, the chance of one player dominating the hand wins remains relatively high, keeping the game duration close to the total number of cards. As the number of players increases, this effect becomes more evenly distributed across the group, and the minimum game duration tends to increase. These findings are further supported by the analysis of hand-winning probabilities discussed in Section~\ref{sec_hand_win_prob}.

On the other hand, Figure~\ref{fig_min_max}(a) displays the maximum number of turns observed in the simulated games. Due to the reshuffling of cards following each hand win, the game inherently avoids non-terminating sequences, ensuring that all games eventually conclude. In our simulations—comprising $10^6$ iterations for each of the 25 combinations of players and decks—all games terminated within a finite number of turns. As the number of decks increases, the maximum number of turns tends to rise accordingly.  The maximum number of turns observed was $26322$, occurring in a configuration with 4 players and 5 decks. Notably, this setup does not represent an extreme case (i.e., it does not involve the maximum number of players). A similar trend is observed with $K=4$, prompting an interesting question: does the longest game duration typically arise from an optimal number of players for a given number of decks, or is this result merely a consequence of limited sampling relative to the low probability of exceedingly long games?

\subsection{Typical game length}
\begin{figure*}
\centering
\includegraphics[width=\textwidth]{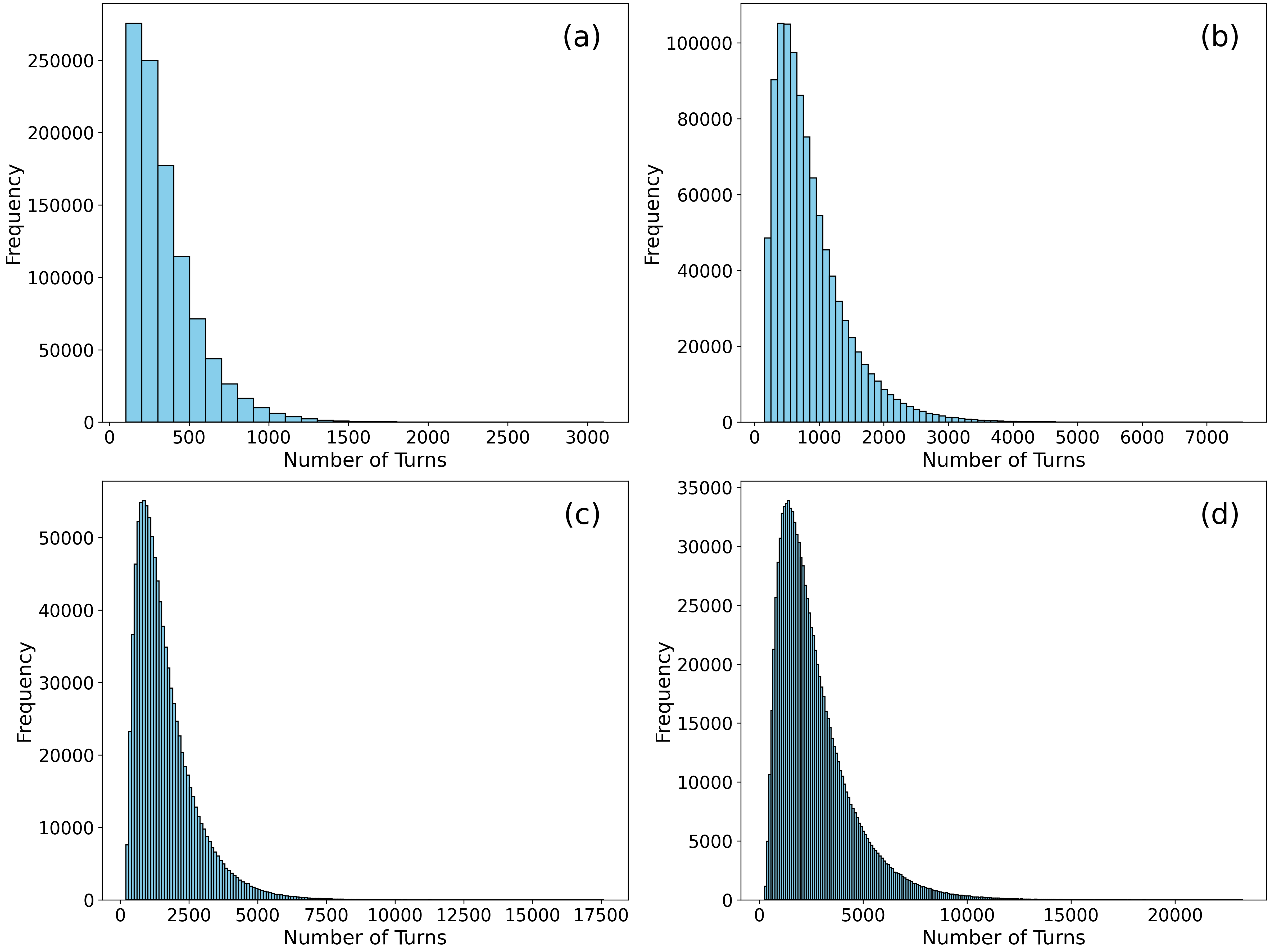}
\caption{ Distribution of number of turns in $10^6$ games for number of players $N$ \& number of decks $K$ for (a)$N=3$ \& $D=2$, (b)$N=3$ \& $D=3$, (c)$N=3$ \& $D=4$,(d)$N=3$ \& $D=5$. Each bin width (x-axis) corresponds to 100 turns. The y-axis shows the count of games that lasted within the range of 100 turns out of all the simulated games for a given $N$ and $K$. \label{fig_turn_dist}}
\end{figure*}

The extreme cases discussed earlier—specifically, the maximum game durations—represent rare events. A natural question that arises is: how long does a typical game usually last? In other words, what is the distribution of game durations? Figure~\ref{fig_turn_dist} illustrates the distribution of the number of turns for four representative cases with $N$ fixed and increasing $K$. It is evident that these distributions peak at significantly lower values than the corresponding maximum durations shown in Figure~\ref{fig_min_max}. The heavy-tailed nature of the distributions highlights that games approaching the maximum number of turns are indeed rare occurrences. However, the peak shifts to higher and higher values as the number of decks is increased, suggesting that those games last longer on average. A random game is expected to last for a duration randomly drawn from this distribution. In this study, we do not characterise these distributions. 

\subsection{Hand size distribution}
\begin{figure*}
\centering
\includegraphics[width=\textwidth]{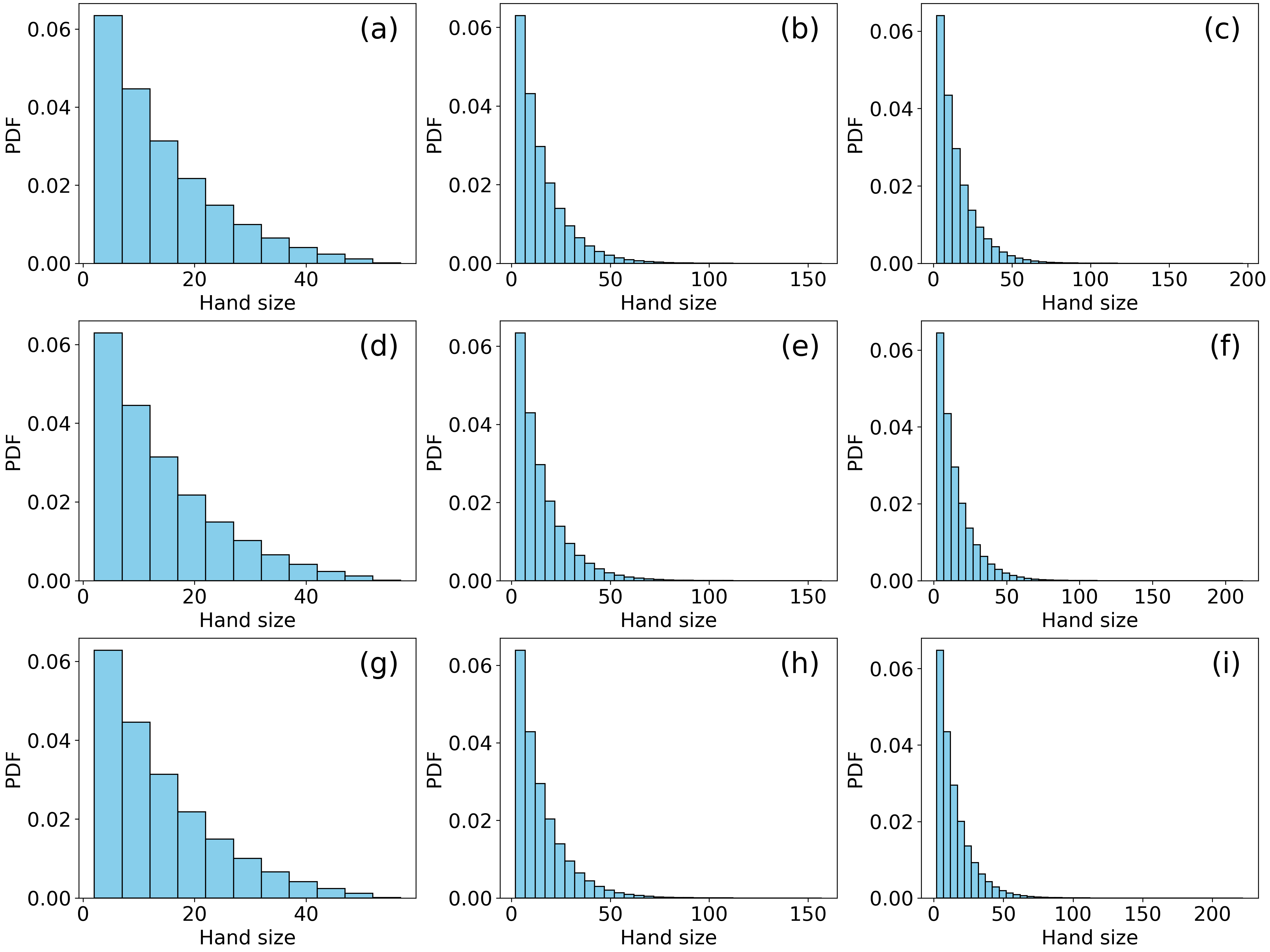}
\caption{ Probability distribution function (PDF) of hand sizes won during the simulation. The x-axis shows the hand sizes (bin width corresponds to 5 cards), and the y-axis shows the PDF of the hand size. Here, only nine representative cases are shown from the total of twenty-five combinations. They are as: (a) $N=3$ \& $K=1$, (b) $N=3$ \& $K=3$, (c) $N=3$ \& $K=5$, (d) $N=4$ \& $K=1$, (e) $N=4$ \& $K=3$, (f) $N=4$ \& $K=5$, (g) $N=5$ \& $K=1$, (h) $N=5$ \& $K=3$, (i) $N=5$ \& $K=5$. \label{fig_hand_size}}
\end{figure*}

We record the number of cards in every hand won during the simulation for each combination of $N$ and $K$. Figure~\ref{fig_hand_size} presents the probability distribution functions (PDFs) of the hand sizes for nine representative cases. The other cases, not shown here, show similar trends. For the sake of comparison, all bin sizes are set to five cards in Fig.~\ref{fig_hand_size}. In each case, the distribution peaks at a number of fewer than five cards, meaning most hands are of size less than 5 cards. As the number of decks increases, the distributions develop heavier tails, indicating a very small but non-zero probability of observing exceptionally large hands. This trend persists regardless of the number of players involved in the game.

\subsection{Player hand win probability}
\label{sec_hand_win_prob}

\begin{figure*}
\centering
\includegraphics[width=\textwidth]{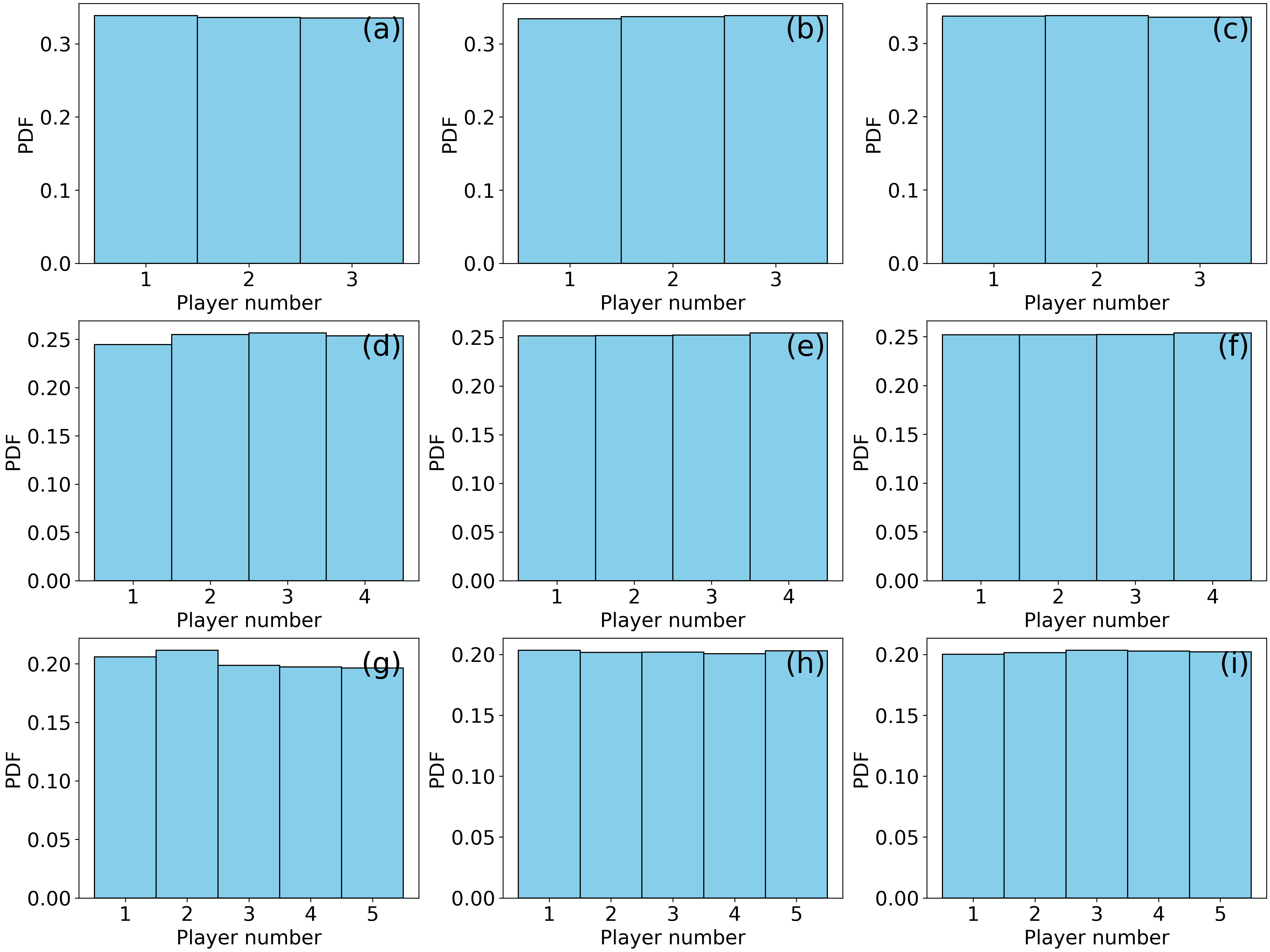}
\caption{ Probability distribution function (PDF) of hands won by individual players. The x-axis shows the player number, and the y-axis shows the PDF of the number of hands won by the player. Here, we show PDF for (a) $N=3$ \& $K=1$, (b) $N=3$ \& $K=3$, (c) $N=3$ \& $K=5$, (d) $N=4$ \& $K=1$, (e) $N=4$ \& $K=3$, (f) $N=4$ \& $K=5$, (g) $N=5$ \& $K=1$, (h) $N=5$ \& $K=3$, (i) $N=5$ \& $K=5$. \label{fig_hand_win}}
\end{figure*}

Here, we answer the question, Who wins the most hands? We present the probability distribution of the number of hands won by individual players. Notably, in a randomly played game, the player who wins the most hands does not necessarily emerge as the overall winner. Victory is determined by the total number of cards accumulated, which may result from capturing a few large hands or numerous smaller ones.

In contrast, if no player wins a hand or if a single player wins all the hands in a game, the remaining players are unable to increase their card count. This scenario leads to the shortest possible game, consisting of turns equal to the total number of cards in play.

As shown in Fig.~\ref{fig_hand_win}, when the game is played with fewer decks, the probability of winning individual hands is distributed slightly unevenly among players. However, this distribution becomes more uniform as the number of decks increases. In games with a smaller number of decks, there is a possibility that one player may have a slight advantage in winning a total number of hands across several games. However, as the number of decks increases, this advantage diminishes, and no player exhibits a statistically higher likelihood of winning a hand.

\subsection{Player game win probability}

\begin{figure*}
\centering
\includegraphics[width=\textwidth]{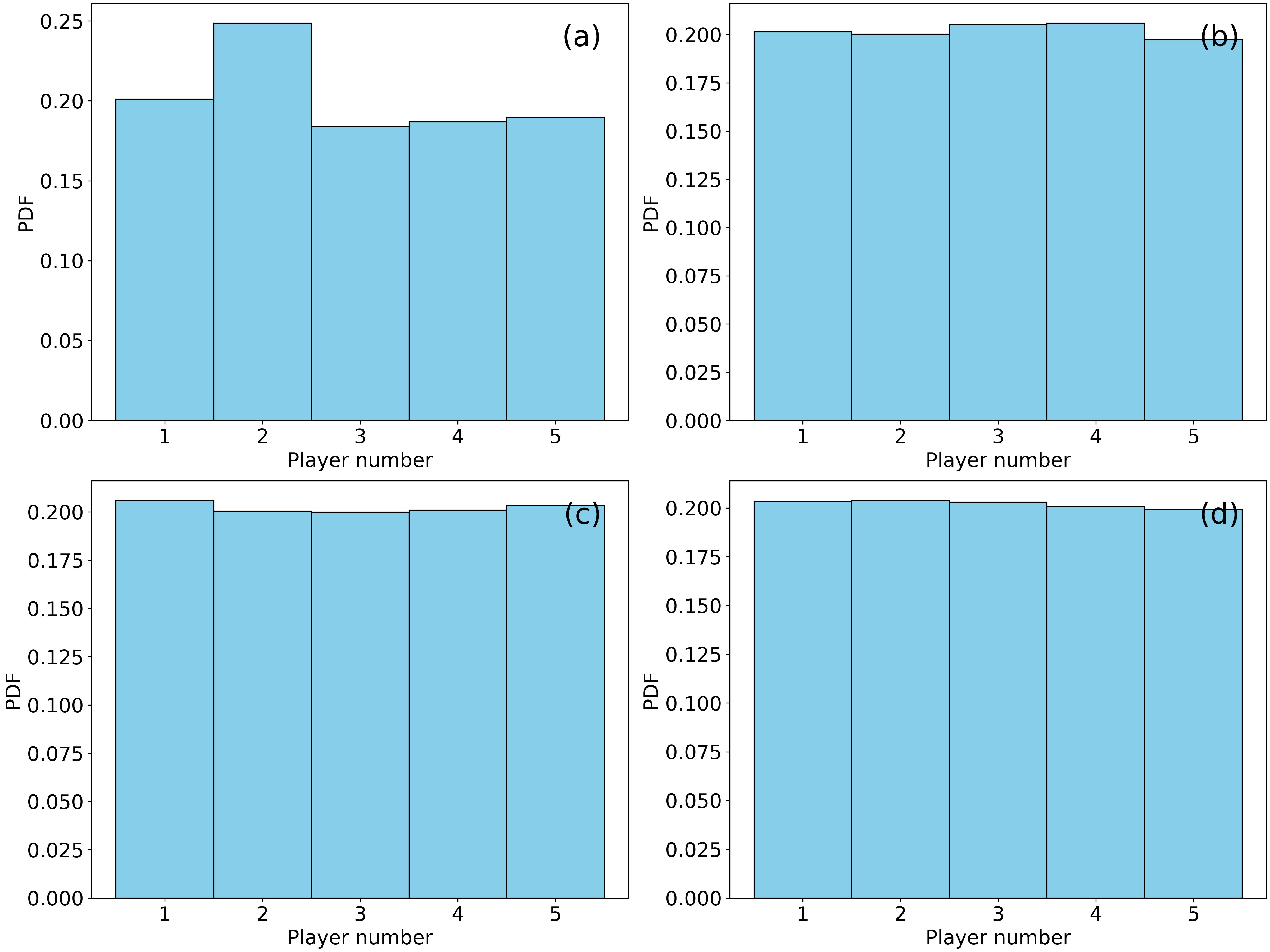}
\caption{ Probability distribution function (PDF) of games won by individual players. The x-axis shows the player number (note - player 1 always starts the game), and the y-axis shows the PDF of the games won by the player. Here, we show the PDF for $N=5$ and $K \in (a) 1, (b) 2, (c) 3, (d) 4$. \label{fig_pdf_win}}
\end{figure*}

In Fig.~\ref{fig_pdf_win}, we present the probability distribution function (PDF) of games won by individual players. In our setup, Player 1 always initiates the game; however, in practice, the starting player could rotate among all participants. As observed in Fig.~\ref{fig_pdf_win}(a), the win probability is slightly uneven when the number of decks is small. Nevertheless, no player position consistently gains a significant advantage under these conditions. For other combinations of the number of players $N$ and decks $K$ (not shown here), a different player may exhibit a slightly higher win probability. As the number of decks increases, the win probability tends to distribute more evenly across all players, indicating that the winner becomes effectively random in a randomly played game.  

\section{Conclusion}
\label{sec:conclusions}
In this study, we conducted an extensive Monte Carlo simulation of the \textit{Bhikar-Sawkar} card game to investigate its statistical properties, including game durations and win probabilities, across a comprehensive range of player and deck configurations. While the game bears structural resemblance to \textit{Beggar-my-neighbour}, subtle differences in rules render \textit{Bhikar-Sawkar} a fundamentally non-deterministic game. These nuances, particularly the intermittent reshuffling of cards, prevent non-terminating configurations and ensure that every game reaches a finite conclusion. However, they also introduce historical dependence on the cards played so far and stochastic complexity into the gameplay, thereby complicating analytical approaches and making theoretical predictions elusive.

Our simulations reveal that both the maximum and minimum game durations are strongly influenced by the number of players and the number of decks. Games with fewer decks tend to be shorter and more prone to dominance by a single player, whereas configurations with larger decks yield longer, more balanced games. We observed that for a given number of decks, there exists an optimal number of players that tends to yield longer game durations. However, this observation warrants further investigation to establish firm conclusions and to determine whether this trend reflects an underlying structural property of the game or arises from statistical fluctuations. 

Importantly, the maximum durations observed in our simulations should not be interpreted as absolute upper bounds; rather, they reflect rare but possible outcomes within a heavy-tailed distribution of game lengths. In contrast, the majority of games terminate significantly earlier, as indicated by the peaked structure of the duration distribution at shorter turn counts. This highlights the stochastic nature of the gameplay and quantifies the likelihood of both typical and extreme outcomes across a wide range of configurations.

Additionally, we observe that in games with a small number of decks, certain players may exhibit a marginal statistical advantage in winning. However, as the number of decks increases, this asymmetry diminishes, and win probabilities become more uniformly distributed among players. This indicates that the finite-size effects associated with smaller deck configurations wane in larger, more randomised systems, reinforcing the inherently random character of the game under such conditions.

This study employs Monte Carlo methods to derive statistical insights into an interesting nondeterministic game and poses the question of whether these probabilities can be predicted using analytical models.
\section*{Acknowledgements} \label{sec:acknowledgements}

We gratefully acknowledge funding by the European Union (EU) under the Horizon Europe research and innovation programme, EIC Pathfinder - grant No. 101187428 (iNSIGHT). The present work was conducted independently and is not related to the objectives or activities of the funded project.

\bibliography{mybib}

\begin{thebibliography}{7}%
\makeatletter
\providecommand \@ifxundefined [1]{%
 \@ifx{#1\undefined}
}%
\providecommand \@ifnum [1]{%
 \ifnum #1\expandafter \@firstoftwo
 \else \expandafter \@secondoftwo
 \fi
}%
\providecommand \@ifx [1]{%
 \ifx #1\expandafter \@firstoftwo
 \else \expandafter \@secondoftwo
 \fi
}%
\providecommand \natexlab [1]{#1}%
\providecommand \enquote  [1]{``#1''}%
\providecommand \bibnamefont  [1]{#1}%
\providecommand \bibfnamefont [1]{#1}%
\providecommand \citenamefont [1]{#1}%
\providecommand \href@noop [0]{\@secondoftwo}%
\providecommand \href [0]{\begingroup \@sanitize@url \@href}%
\providecommand \@href[1]{\@@startlink{#1}\@@href}%
\providecommand \@@href[1]{\endgroup#1\@@endlink}%
\providecommand \@sanitize@url [0]{\catcode `\\12\catcode `\$12\catcode
  `\&12\catcode `\#12\catcode `\^12\catcode `\_12\catcode `\%12\relax}%
\providecommand \@@startlink[1]{}%
\providecommand \@@endlink[0]{}%
\providecommand \url  [0]{\begingroup\@sanitize@url \@url }%
\providecommand \@url [1]{\endgroup\@href {#1}{\urlprefix }}%
\providecommand \urlprefix  [0]{URL }%
\providecommand \Eprint [0]{\href }%
\providecommand \doibase [0]{http://dx.doi.org/}%
\providecommand \selectlanguage [0]{\@gobble}%
\providecommand \bibinfo  [0]{\@secondoftwo}%
\providecommand \bibfield  [0]{\@secondoftwo}%
\providecommand \translation [1]{[#1]}%
\providecommand \BibitemOpen [0]{}%
\providecommand \bibitemStop [0]{}%
\providecommand \bibitemNoStop [0]{.\EOS\space}%
\providecommand \EOS [0]{\spacefactor3000\relax}%
\providecommand \BibitemShut  [1]{\csname bibitem#1\endcsname}%
\let\auto@bib@innerbib\@empty
\bibitem [{\citenamefont {Sharna}\ \emph {et~al.}(2021)\citenamefont {Sharna},
  \citenamefont {Sharma}, \citenamefont {Doyle}, \citenamefont {Marcelo},\ and\
  \citenamefont {Kumar}}]{Sharna_Sharma_Doyle_Marcelo_Kumar_2021}%
  \BibitemOpen
  \bibfield  {author} {\bibinfo {author} {\bibfnamefont {S.}~\bibnamefont
  {Sharna}}, \bibinfo {author} {\bibfnamefont {S.}~\bibnamefont {Sharma}},
  \bibinfo {author} {\bibfnamefont {P.}~\bibnamefont {Doyle}}, \bibinfo
  {author} {\bibfnamefont {L.}~\bibnamefont {Marcelo}}, \ and\ \bibinfo
  {author} {\bibfnamefont {D.}~\bibnamefont {Kumar}},\ }\href {\doibase
  10.15663/wje.v26i2.881} {\bibfield  {journal} {\bibinfo  {journal} {Waikato
  Journal of Education}\ }\textbf {\bibinfo {volume} {26}},\ \bibinfo {pages}
  {51–64} (\bibinfo {year} {2021})}\BibitemShut {NoStop}%
\bibitem [{\citenamefont {P.Eng.}(2016)}]{asee_peer_27135}%
  \BibitemOpen
  \bibfield  {author} {\bibinfo {author} {\bibfnamefont {R.~A.~B.}\
  \bibnamefont {P.Eng.}},\ }in\ \href@noop {} {\emph {\bibinfo {booktitle}
  {2016 ASEE Annual Conference \& Exposition}}},\ \bibinfo {series and number}
  {\bibinfo {number} {10.18260/p.27135}}\ (\bibinfo  {publisher} {ASEE
  Conferences},\ \bibinfo {address} {New Orleans, Louisiana},\ \bibinfo {year}
  {2016})\ \bibinfo {note} {https://peer.asee.org/27135}\BibitemShut {NoStop}%
\bibitem [{\citenamefont {Madsen}\ \emph {et~al.}(1998)\citenamefont {Madsen},
  \citenamefont {Nielsen},\ and\ \citenamefont {Jensen}}]{madsen}%
  \BibitemOpen
  \bibfield  {author} {\bibinfo {author} {\bibfnamefont {A.}~\bibnamefont
  {Madsen}}, \bibinfo {author} {\bibfnamefont {L.}~\bibnamefont {Nielsen}}, \
  and\ \bibinfo {author} {\bibfnamefont {F.}~\bibnamefont {Jensen}},\ }in\
  \href@noop {} {{\selectlanguage {English}\emph {\bibinfo {booktitle}
  {Proceedings of the 11th International Florida Artificial Intelligence
  Research Symposium Conference}}}},\ \bibinfo {editor} {edited by\ \bibinfo
  {editor} {\bibnamefont {{Cook, Diane J. (ed.)}}}}\ (\bibinfo  {publisher}
  {AAAI Press},\ \bibinfo {address} {United States},\ \bibinfo {year} {1998})\
  pp.\ \bibinfo {pages} {435--439},\ \bibinfo {note} {probSy - A System for the
  Calculation of Probabilities in the Card Game Bridge ; Conference date:
  19-05-2010}\BibitemShut {NoStop}%
\bibitem [{\citenamefont {Paulhus}(1999)}]{Paulhus01021999}%
  \BibitemOpen
  \bibfield  {author} {\bibinfo {author} {\bibfnamefont {M.~M.}\ \bibnamefont
  {Paulhus}},\ }\href {\doibase 10.1080/00029890.1999.12005024} {\bibfield
  {journal} {\bibinfo  {journal} {The American Mathematical Monthly}\ }\textbf
  {\bibinfo {volume} {106}},\ \bibinfo {pages} {162} (\bibinfo {year}
  {1999})},\ \Eprint
  {http://arxiv.org/abs/https://doi.org/10.1080/00029890.1999.12005024}
  {https://doi.org/10.1080/00029890.1999.12005024} \BibitemShut {NoStop}%
\bibitem [{\citenamefont {Casella}\ \emph {et~al.}(2024)\citenamefont
  {Casella}, \citenamefont {Anderson}, \citenamefont {Kleber}, \citenamefont
  {Mann}, \citenamefont {Nessler}, \citenamefont {Rucklidge}, \citenamefont
  {Williams},\ and\ \citenamefont {Wu}}]{casella2024}%
  \BibitemOpen
  \bibfield  {author} {\bibinfo {author} {\bibfnamefont {B.}~\bibnamefont
  {Casella}}, \bibinfo {author} {\bibfnamefont {P.~M.}\ \bibnamefont
  {Anderson}}, \bibinfo {author} {\bibfnamefont {M.}~\bibnamefont {Kleber}},
  \bibinfo {author} {\bibfnamefont {R.~P.}\ \bibnamefont {Mann}}, \bibinfo
  {author} {\bibfnamefont {R.}~\bibnamefont {Nessler}}, \bibinfo {author}
  {\bibfnamefont {W.}~\bibnamefont {Rucklidge}}, \bibinfo {author}
  {\bibfnamefont {S.~G.}\ \bibnamefont {Williams}}, \ and\ \bibinfo {author}
  {\bibfnamefont {N.}~\bibnamefont {Wu}},\ }\href
  {https://arxiv.org/abs/2403.13855} {\enquote {\bibinfo {title} {A
  non-terminating game of beggar-my-neighbor},}\ } (\bibinfo {year} {2024}),\
  \Eprint {http://arxiv.org/abs/2403.13855} {arXiv:2403.13855 [math.CO]}
  \BibitemShut {NoStop}%
\bibitem [{\citenamefont {Lakshtanov}\ and\ \citenamefont
  {Aleksenko}(2013)}]{Lakshtanov2013}%
  \BibitemOpen
  \bibfield  {author} {\bibinfo {author} {\bibfnamefont {E.~L.}\ \bibnamefont
  {Lakshtanov}}\ and\ \bibinfo {author} {\bibfnamefont {A.~I.}\ \bibnamefont
  {Aleksenko}},\ }\href {\doibase 10.1134/S0032946013020051} {\bibfield
  {journal} {\bibinfo  {journal} {Problems of Information Transmission}\
  }\textbf {\bibinfo {volume} {49}},\ \bibinfo {pages} {163} (\bibinfo {year}
  {2013})}\BibitemShut {NoStop}%
\bibitem [{\citenamefont {Durve}(2025)}]{code_link}%
  \BibitemOpen
  \bibfield  {author} {\bibinfo {author} {\bibfnamefont {M.}~\bibnamefont
  {Durve}},\ }\href {\doibase 10.5281/zenodo.15351816} {\enquote {\bibinfo
  {title} {Monte carlo simulation code for bhikar-sawkar card game},}\ }
  (\bibinfo {year} {2025})\BibitemShut {NoStop}%
\end{thebibliography}%

\end{document}